\documentclass[%
reprint,
onecolumn,
 amsmath,amssymb,
 aps,
]{revtex4-2}

\usepackage{graphicx}
\usepackage[hidelinks]{hyperref}

\usepackage{dcolumn}
\usepackage{bm}
\usepackage{amsmath}
\usepackage{makecell}
\usepackage{color}
\usepackage{ulem}

\usepackage{comment}

\begin{document}


\title[Enstrophy Production in the Near-Wall of a Turbulent Channel Flow]{The Role of Normal and Non-Normal Contributions to Enstrophy Production in the Near-Wall Region of a Turbulent Channel Flow}

\author{C.~J.~Keylock${}^{1}$}
\email{C.J.Keylock@lboro.ac.uk}

\affiliation{${}^{1}$Loughborough University, School of Architecture, Building and Civil Engineering, Loughborough, Leicestershire, U.K.}

\date{\today}

\begin{abstract}
The turbulent boundary-layer is a region where both preferential dissipation of energy and the production of significant vorticity arises as a consequence of the strong velocity gradients. Previous work has shown that, following a Reynolds decomposition of the enstrophy production, the purely fluctuating contribution is the dominant term and that near the wall this varies in a complex manner with height. In this study we additionally decompose the strain rate and vorticity terms into normal and non-normal components using a Schur decomposition and are able to explain all these features in terms of contributions at different heights from constituents involving different combinations of normal and non-normal quantities. What is surprising about our results is that while the mean shear and the action of larger scale structures should mean that non-normal effects are of over-riding importance, the most important individual term involves the fluctuating, normal straining in the transverse direction. Furthermore, the reason that the term that involves only non-normal contributions is smaller on average than that involving normal straining coupled to non-normal vorticity is that in the former case there are individual constituents that are negative in the mean. Hence, we not only explain the nature of near-wall enstrophy production in greater detail, but highlight how local straining that is orthogonal to the direction of the dominant mean and fluctuating shear plays a crucial role in amplifying vorticity that is yet to have developed sufficiently to gain a solid body rotational component. 
\end{abstract}

\maketitle

\section{Introduction}
The fundamental result underpinning much of our understanding of the multi-scale behaviour of turbulence from a statistical perspective is due to \cite{K41a,K41b}. With a sufficiently large separation between the scales at which energy is injected and those where it is dissipated, we have that
\begin{equation}
    \langle \delta u_{r}^{3}\rangle - 6\, \nu \frac{d} {d\,r} \langle \delta u_{r}^{2}\rangle = -\frac{4}{5}\langle\epsilon\rangle r,
    \label{eq.Kolm}
\end{equation}
where $\langle\ldots\rangle$ is an ensemble average, $\delta u_{r}$ is the velocity difference for the longitudinal component at a separation $r$, $\nu$ is the kinematic viscosity, and $\epsilon$ is the turbulence dissipation rate per unit mass. However, moving beyond a simple statistical description has proven difficult due to the suite of vortices at different scales with varying topologies that generate a local physics that may depart significantly from the average picture. For example, see \cite{ishihara09} for a review of numerical work examining (\ref{eq.Kolm}) for isotropic turbulence as a function of Reynolds number, including the distinctly non-Gaussian dissipation and enstrophy statistics that result from vortical behaviour and \cite{wilczek24} for a review of recent work on the multi-scale behaviour of the velocity gradient tensor, from which the straining and enstrophy may be derived. 

When we attempt to translate these concepts from isotropic turbulence to boundary-layers and channel flows, the key difference is that the presence of the wall provides a means to generate and sustain turbulent motions \cite{lumley64}. As a consequence, the generalization of (\ref{eq.Kolm}) due to \cite{vkh38}, \cite{MY75} and \cite{hill02}, which retains information on local spatial fluxes and the local time derivative has been used in several studies to gain a deeper insight into these processes \cite{danaila01,marati04, cimarelli13,cimarelli16}. Such work has provided a term-by-term decomposition of the flow, providing valuable insight into the dynamics. In particular, \cite{cimarelli16} identified three driving mechanisms using this approach, with those associated with attached and detached flow structures distinguished by a threshold height for the centre of the vortices found to lie at 20 wall units in the wall-normal ($y$) direction. 

The dynamics of these near-wall vortices is strongly associated with the attached eddy hypothesis \cite{Townsend76}, and a great deal of recent work has demonstrated the general correctness of this formalism. For example, \cite{jimenez08} were able to show that the squared velocity intensities scaled logarithmically away from the wall as predicted in the original theory and \cite{hwang15} determined a self-similarity in the spanwise length of self-sustaining motions, which could be linked explicitly to the structures postulated in the attached eddy hypothesis. In addition, evidence has also emerged that the larger scales in a channel flow can maintain themselves rather than being the consequence of an amalgamations of smaller scale streaks or hairpin vortices \cite{hwang10}. This self-sustaining process is similar to that previously identified by \cite{hamilton95} and \cite{schoppa02} in the near-wall region for the smallest attached eddies. \cite{hamilton95} initiated a Couette flow numerically with just the vortices present and any cross-flow fixed (to prevent vortex decay) and then looked for the development of streaks. It was found that the streamwise vortices acting on
the mean flow produced the streaks by momentum redistribution. However, breakdown was a consequence of the instability of the streaks rather than the mean flow or the streamwise vortices, with vortex regeneration a consequence of nonlinear interactions. A critical threshold for vortex circulation needed to be maintained for this cycle to be self-sustaining, with the circulation peaking during breakdown and then decaying, with streaks forming during the periods of vortex decay.  

In terms of the scale-by-scale organization of flow structures and the development of an energy spectrum, \cite{goto08} proposed that homogeneous, isotropic turbulence (HIT) was organized such that smaller-scale vortex tubes form in the regions of high straining between larger, anti-parallel tubular structures and wrap around them in a perpendicular orientation. Extending this work to wall-bounded flows, \cite{motoori19} found that vortices sufficiently small relative to the distance from the wall were formed by stretching due to the action of vortices twice as large. This is similar to the observation by \cite{cardesa17} that in the turbulence cascade, packets of energetic motions of a given scale, first appear within packets with double this scale, and dissipate within those that are half as large. In contrast, and related to the attached eddy hypothesis, \cite{motoori19} found that the vortices scaling on the distance from the wall developed as a consequence of the mean gradients in the flow field rather than this non-local stretching mechanism. The anti-parallel arrangement of vortices as a function of the scale of the cascade has also been detected in vortex ring collision experiments \cite{mckeown20}. Those authors found that, at a given scale, pairs of anti-parallel vortices interact according to the Biot-Savart law to generate a subsequent flow structure via the elliptical instability \cite{kerswell02} and proposed that this mechanism was analogous to that identified by \cite{goto08}.

The manner by which the vortices develop and the flow dissipates energy may be considered from the perspective of enstrophy production and strain production, the terms contributing to the third invariant of the velocity gradient tensor (VGT) \cite{carbone22}. \cite{taylor37} observed that enstrophy production is positive in the mean and hypothesised that vortex stretching is the principal means by which turbulent flows dissipate energy. However, further work by \cite{betchov56} showed that, in fact, vortex compression is critical. Betchov used an expression due to \cite{townsend51} for the relation between average strain production and average enstrophy production in the frame of the principal axes of the straining field to illustrate this:
\begin{equation}
\label{eq.Towns}
  -\langle e_{1}e_{2}e_{3}\rangle = \langle e_{1}\Omega_{1}^{2} + e_{2}\Omega_{2}^{2} + e_{3}\Omega_{3}^{2}\rangle.
\end{equation}
Here, the $e_{i}$ are the eigenvalues of the strain rate tensor (ordered in descending magnitude in this instance) and the $\Omega_{i}$ are the rotations about an axis parallel to the principal axes of the straining field. Incompressibility means that $e_{2}$ and $e_{3}$ must have the same sign and this is opposite to $e_{1}$. Vortex stretching corresponds to $e_{1} > 0$, but Betchov noted that for both sides of \ref{eq.Towns} to be positive requires $e_{1} < 0$ given that $\Omega_{1}$ is attenuated but $\Omega_{2}$ and $\Omega_{3}$ are enhanced in such a compressive, ``jet collision'' regime.

Greater understanding of this result stems from the work by \cite{jimenez93} who showed that enstrophy production is correlated to both the enstrophy and the total strain, but more strongly to the latter, and from the work of \cite{tsinober01} who showed that strain production is also associated with the total strain, but exhibits very little relation to the enstrophy. The reason for this complexity is that the non-local nature of a turbulent flow field as a consequence of the action of the pressure terms \cite{ohkitani95} means that any given point may have high values for strain production and small values for enstrophy production despite the fact that on average they are in balance.  

The enstrophy production in the flow is given by $\omega_{i}^{A}S_{ij}^{A}\omega_{j}^{A}$ where $\bm{A}$ is the velocity gradient tensor, 
$A_{ij} = \partial u_{i} / \partial x_{j}$, $S_{ij}$ is the strain rate, and $\omega_{i}$ is the vorticity vector, $\omega_{i} = -\epsilon_{ijk}\Omega_{jk}$, where $\epsilon_{ijk}$ is the Levi-Civita symbol and
\begin{align}
\label{eq.strain}
\bm{S}_{A} &= \frac{1}{2}\left(\bm{A} + \bm{A}^{*}\right),\\
\bm{\Omega}_{A} &= \frac{1}{2}\left(\bm{A} - \bm{A}^{*}\right),
\label{eq.rot}
\end{align}
are the strain and rotation tensors, respectively. 

One way to isolate the non-normal contributions to the flow dynamics has been to decompose $\bm{A}$ using a Schur decomposition \cite{k18,k19}. This framework has been applied to study turbulence close to interfaces \cite{boukharfane21}, in the near-wake region of spatially developing flows \cite{beaumard19} and near the wall \cite{k22}. It provides a means to understand how normal and non-normal processes affect the flow dynamics. For example, \cite{k18} showed that the preferential alignment between the vorticity vector and the eigenvector for $e_{2}$ \cite{kerr85,ashurst87} arose in HIT in two different ways in the regions that dominated the generation of significant alignments:
   Where the third invariant of the VGT, $R$, was positive and the second invariant, $Q$, was negative, the vorticity orientation was dictated by the non-normal contribution and there was also a strong mutual alignment between the eigenvectors for the intermediate strain eigenvalues for both the normal and non-normal straining fields. In contrast,  where $Q > 0$ and $R< 0$ the local vorticity dominates in determining the vorticity orientation and the most extensive eigenvalue for the local straining is aligned with the intermediate eigenvalue for the non-normal straining. 

Work on using the Schur decomposition to distinguish normal and non-normal dynamics explicitly has also begun to feed into engineering model development. For example, \cite{Yu21} used an index of the relative magnitude of the normal and non-normal parts of the VGT from \cite{k18} to show that conditioning the subgrid scale energy and enstrophy dissipation on the sign of this index is a useful means to build a sub-grid closure for large-eddy simulations of compressible flow reflecting the varying nature of these terms near the wall. Given the utility of this approach, gaining an understanding of some of the above results for boundary-layers or channel flows from the perspective of the normal and non-normal dynamics would appear to be potentially insightful. In this study we focus upon the enstrophy production near the wall and, thus, the manner by which turbulence is generated in near-wall regions. In order to achieve this we first outline the Schur decomposition approach before deriving expressions for the enstrophy production in terms of the mean and fluctuating quantities \cite{motoori19} \textit{as well as} the normal and non-normal contributions. We then use a direct numerical simulation of a turbulent channel flow with a shear velocity Reynolds number of $\sim 1000$ \cite{yili,graham16} to establish which terms drive the dynamics of a near-wall channel flow. 

\section{Normal and Non-Normal Contributions to Velocity Gradient Tensor Dynamics}
In this study we adopt an approach to an analysis of the VGT based on the Schur decomposition  of this tensor \cite{schur1909}. Such a transform was adopted by \cite{lietal14} to undertake a triple decomposition of the flow field into a rigid body rotation term, a shearing terms and a stretching/compressive term (see also, \cite{kolar13} and \cite{liu22}). The rigid body rotation component then underpins the Liutex definition of a coherent structure \cite{xu19} to provide a definition of coherent flow structures, while \cite{zhu21} has defined a real Schur flow and used this notion to explore, in particular, the dynamics of rotational and compressible flows. 

The motivation for the use of the Schur decomposition here takes as its starting point the restricted Euler model for the dynamics of the VGT \cite{cantwell92}. Taking the spatial derivative of the Navier-Stokes equations gives an evolution equation for the dynamics of $\bm{A}$, for an incompressible fluid  where $\mbox{tr}(\bm{A}) = 0$ and $\mbox{tr}(\ldots)$ is the trace:
\begin{align}
\frac{\partial \bm{A}}{\partial t} + (\bm{u} \cdot \nabla)\bm{A} = -\left(\bm{A}^2-\frac{\mbox{tr}(\bm{A}^2)}{3}\bm{I}\right) - \bm{H} + \nu \nabla^2 \bm{A},
\label{eq.NSgrad}
\end{align}
where $\bm{H}$ is the anisotropic/deviatoric part of the pressure Hessian, $\bm{H}\equiv \nabla\nabla p -\nabla^2p\bm{I}/3$, and $\bm{I}$ is the identity matrix. \cite{cantwell92} simplified (\ref{eq.NSgrad}) by neglecting the viscous term and the deviatoric part of the pressure Hessian to give two coupled ordinary differential equations for the Lagrangian evolution of the VGT:
\begin{align}
\nonumber
\frac{d Q}{d t} &= -3 {R} \\
\frac{d R}{d t} &= \frac{2}{3} {Q}^{2}, 
\label{eq.QR}
\end{align}
where
\begin{align}
\label{eq.Q}
    Q &= (1 - \delta_{ij}) \sum \lambda_{i}\lambda_{j} \equiv \frac{1}{2}\left(\Vert\bm{\Omega}_{A}\Vert^{2} - \Vert\bm{S}_{A}\Vert^{2}\right), \\
    R &= \prod_{i} \lambda_{i} \equiv -\mbox{det}(\bm{S}_{A}) - \mbox{tr}(\bm{\Omega}_{A}^{2}\bm{S}_{A}),
    \label{eq.R}
\end{align}
where $\delta_{ij}$ is Kronecker's delta, $\mbox{det}(\ldots)$ is the determinant, $\Vert\ldots\Vert$ is the Frobenius norm and the $\lambda_{i}$ are the eigenvalues of $\bm{A}$. Given that $Q$ and $R$ may be expressed in terms of the eigenvalues of $\bm{A}$ (i.e. the normal part of the tensor), it follows that the remaining dynamics due to the viscous term and the deviatoric part of the pressure Hessian, must include the non-local effects \cite{ohkitani95} and that their action will be reflected in the non-normal parts of $\bm{A}$. This is made explicit by adopting the Schur transform and here we follow \cite{k18} by considering the VGT to be formed additively by two constituents
\begin{equation}
  \bm{A} = \bm{B} + \bm{C}, 
  \label{eq.ABC}
\end{equation} 
where the eigenvalue-related, normal contributions are in $\bm{B}$ and $\bm{C}$ contains the non-normal contribution. The complex Schur decomposition of $\bm{A}$ is
\begin{equation}
\bm{A} = \bm{U} \bm{T} \bm{U}^{*},
\end{equation}
where $\bm{U}$ is unitary i.e. $\bm{U}\bm{U}^{*} = \bm{I}$, the identity matrix, and the asterisk indicates the transpose. The upper triangular tensor $\bm{T} = \bm{\Lambda} + \bm{N}$ consists of a diagonal matrix of eigenvalues, $\bm{\Lambda}$, where $\Lambda_{i,i} = \lambda_{i}$, and the non-normal contribution, $\bm{N}$. Decomposing $\bm{T}$ in this way permits a reconstruction:
\begin{align}
\notag{}
\bm{B} &= \bm{U} \bm{\Lambda} \bm{U}^{*}\\
\bm{C} &= \bm{U} \bm{N} \bm{U}^{*}.
\end{align}
Hence, the relative magnitude of the normal and non-normal constituents may be expressed as
\begin{equation}
  \kappa_{B,C} = \frac{||\bm{B}||-||\bm{C}||}{||\bm{B}||+||\bm{C}||}, 
\end{equation}
and \cite{k18} showed that for homogeneous isotropic turbulence (HIT) at a Taylor Reynolds number of 433, on average $\kappa_{B,C}$ was very close to zero, while for gas-liquid two-phase flows \cite{boukharfane21} found that $\kappa_{B,C}$ became more negative close to the interface between the species. For the spatially developing wake shed from a cylinder and for a developing mixing layer, \cite{beaumard19} found that $\kappa_{B,C}$ was more negative than for HIT, particularly in the strain-dominated parts to the flow where the wake was becoming established. In a channel flow, \cite{k22} showed that $\kappa_{B,C}$ was negative on average for $y^{+} \le 70$. This also held for the various flow quadrants \cite{bt86}, with the exception of the sweeping motions, where the average value was negative in the viscous sub-layer but close to zero for $20 \le y^{+} \le 70$. 

Returning to (\ref{eq.Q}) and (\ref{eq.R}) but focussing on the terms derived from the rotation and straining tensors rather than the eigenvalues, we see that $Q$ is the difference between the enstrophy and the total strain, while $R$ is the difference between the strain production and the enstrophy production. Non-normality contributes equally to each of these terms meaning it is eliminated by the subtraction, which is why $Q$ and $R$ may be written in terms of the $\lambda_{i}$. Explicitly, we have that
\begin{align}
\label{eq.OmA}
    \Vert\bm{\Omega}_{A}\Vert^{2} &= \Vert\bm{\Omega}_{B}\Vert^{2} + \Vert\bm{\Omega}_{C}\Vert^{2}, \\ 
    \Vert\bm{S}_{A}\Vert^{2} &= \Vert\bm{S}_{B}\Vert^{2} + \Vert\bm{\Omega}_{C}\Vert^{2},
    \label{eq.SA}
\end{align}
i.e. the enstrophy may be written in terms of the normal enstrophy and the non-normality, and the total strain rate in terms of the normal strain rate and the non-normality. Similarly, (\ref{eq.R}) may be written as
\begin{align}
\label{eq.strainprod}
-\mbox{det}(\bm{S}_{A}) &= -\mbox{det}(\bm{S}_{B}) -\mbox{det}(\bm{S}_{C}) +\mbox{tr}(\bm{\Omega}_{C}^{2}\bm{S}_{B}) \\
\mbox{tr}(\bm{\Omega}_{A}^{2}\bm{S}_{A}) &= \mbox{tr}(\bm{\Omega}_{B}^{2}\bm{S}_{B}) -\mbox{det}(\bm{S}_{C}) +\mbox{tr}(\bm{\Omega}_{C}^{2}\bm{S}_{B}),
    \label{eq.enstroprod}
\end{align}
where the three terms on the right-hand side of (\ref{eq.strainprod}) are referred to as the normal strain production, non-normal production and interaction production, respectively, and the first term on the right-hand side of (\ref{eq.enstroprod}) is the normal enstrophy production. The two new terms contain the non-normal effects and, at least for HIT, the distribution function for the non-normal production is approximately symmetric while the interaction production is preferentially positive. Hence, it is the latter that drives the general positivity of $-\mbox{det}(\bm{S}_{A})$ and $\mbox{tr}(\bm{\Omega}_{A}^{2}\bm{S}_{A})$ because the sign of $-\mbox{det}(\bm{S}_{B})$ is dictated by the sign of the product of the straining eigenvalues, which is positive for $R > 0$ and negative for $R < 0$, while  $\Vert\bm{\Omega}_{B}\Vert^{2} = 0$, and $\mbox{tr}(\bm{\Omega}_{B}^{2}\bm{S}_{B}) = 0$ are zero if the strain rate tensor eigenvalues are all real or, otherwise, the sign for $\mbox{tr}(\bm{\Omega}_{B}^{2}\bm{S}_{B})$ is opposite to that for $-\mbox{det}(\bm{S}_{B})$. 

The intention in this manuscript is to apply the Schur decomposition framework to the Reynolds decomposition analysis of near-wall enstrophy production by \cite{motoori19} to explain the complex results those authors obtained, such as the changes in sign of the contributing terms with distance from the wall that they discovered. In the next section of this paper we briefly review the simulation used in this study, which is in the public domain, aiding reproducibility of this work. We then study the enstrophy production in detail, with sampling of vertical profiles that are separated in time and space to obtain pseudo-independent samples that are then averaged to obtain converged results. This approach is complemented by the visualisation of the various components of the enstrophy production using the same snapshot of the flow field. 

\section{The Numerical Simulation}
In this study we make use of the Johns Hopkins Turbulence database channel flow simulation  \cite{jhu17,yili}, which is a direct numerical simulation of a
wall-bounded channel flow with periodic boundary conditions in the longitudinal and transverse directions \cite{graham16}. No slip conditions are imposed at the top and bottom walls and the flow was initially driven to maintain a bulk velocity of $U = 1$ until stationarity was achieved. A mean pressure gradient was then applied to give the same shear stress as the previous step. The equations were solved with the PoongBack code \cite{lee13}, with data stored at $2048 \times 512 \times 1536$ spatial positions for a domain size of $L_{x} = 8\pi h$, $L_{y} = 2h$, and $L_{z} = 3\pi h$, with $h$ the channel half height. Thus, the computational cells had a spacing of $x^{+} = 12.26$ and $z^{+} = 6.13$ wall units in the longitudinal and transverse directions, respectively, while the wall-normal resolution varied from $0.02 < y^{+} < 6.16$ from the first node from the wall to the centre-line. The shear velocity Reynolds number was $\mbox{Re}_{\tau} \equiv u_{\tau} h / \nu = 1000$, with a bulk velocity Reynolds number of $\mbox{Re} \equiv U h / \nu = 20 000$ and a kinematic viscosity of $\nu = 5 \times 10^{-5}$ (dimensionless). The best-fit mean velocity profile attained the logarithmic shape conforming to the law-of-the-wall for $30 \lesssim y^{+} \lesssim 250$. The simulation was run for just under 26 dimensionless time steps, with results stored every $\delta t = 0.0065$ steps. Further details of the simulation may be found in \cite{jhu17}. 

 \begin{figure}
 \centering
 \includegraphics[width=0.95\textwidth]{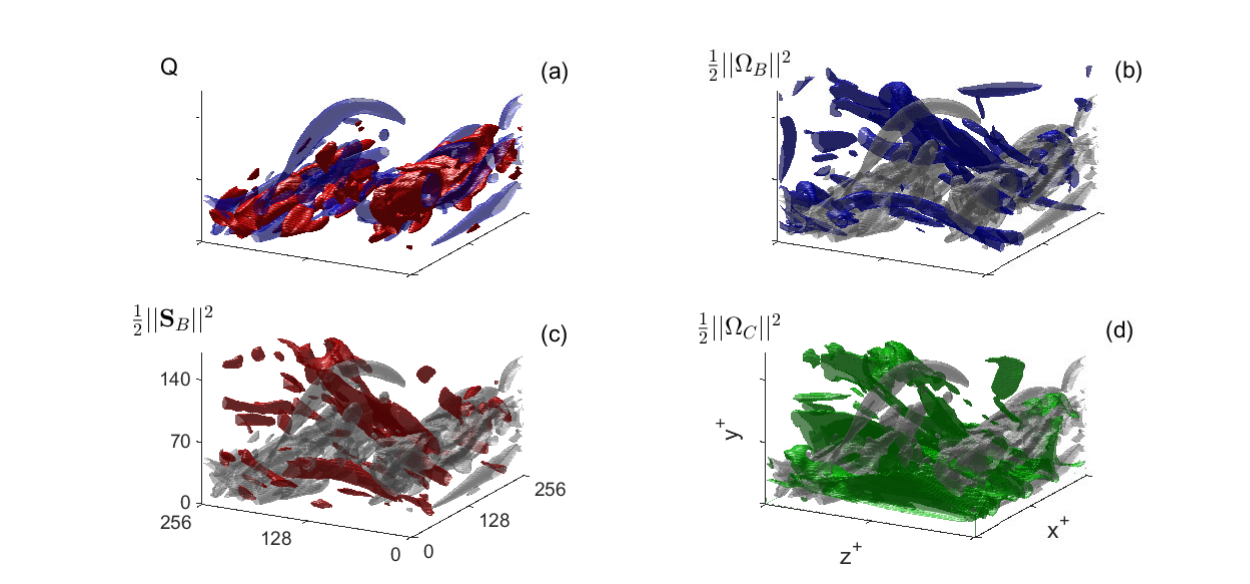}
 \caption{The visualised flow structures for a $x^{+} = 256$ by $y^{+} = 170$ by $z^{+} = 256$ domain are shown in (a) with positive $Q$ in blue and negative in red. The remaining panels show all large $|Q|$ regions in grey and then highlight the individual terms in (\ref{eq.OmA}) and (\ref{eq.SA}), with $\Vert\bm{\Omega}_{B}\Vert^{2}$ in (b), $\Vert\bm{S}_{B}\Vert^{2}$ in (c), and $\Vert\bm{\Omega}_{C}\Vert^{2}$ in (d). Values are made dimensionless using $(\frac{1}{2}\langle \Vert\bm{\Omega}_{A}\Vert^{2} \rangle)^{\frac{3}{2}}$.}
 \label{fig.QR}
  \end{figure}

In order to generate the various vertical profiles of flow quantities shown in this paper we divided the $0 < y^{+} < 70$ range of heights into 128 planes of equal vertical separation. Then at time intervals of $\Delta t \in \{0,2,\ldots,26\}$, we extracted points at regular increments in each horizontal plane such that $256 \times 256$ points were extracted within a $\pi \times \pi$ region, which corresponds to the mesh resolution in the downstream direction. Hence, points were spaced apart in both time and space so that there was no significant correlation between neighbours.  

In contrast, for the visualizations, such as in Fig. \ref{fig.QR}, we isolated a region that was  $x^{+} = 256$ by $y^{+} = 170$ by $z^{+} = 256$, with data extracted every wall unit in each direction. 
An additional criterion was used to remove blocks of contiguous points smaller than a volume of 43 cubic wall units (i.e. a cube with sides of 3.5 wall units). Figure \ref{fig.QR}a shows the selected region in terms of the threshold for $Q$, with $Q > 4\overline{|Q|}$ shown in blue and $Q < -4\overline{|Q|}$ in red. The other three panels 
show the $|Q| > 4\overline{|Q|}$ regions in grey and the locations where the normal enstrophy, normal strain rate and non-normality exceed $6\overline{|Q|}$ in blue, red, and green, respectively. Features that can be seen here include, in panel (a), longitudinally oriented, larger diameter strain-dominated structures at the wall, with positive $Q$ structures wrapped around them. Panels (b) and (c) show a spatial association between regions with strong normal enstrophy and normal straining and these  have a predominant transverse orientation. The non-normality shown in green in panel (d) exhibits a particularly strong concentration in the near-wall region. 

\section{Enstrophy Production}
\label{sect.enstroprod}
 \begin{figure}
 \centering
 \includegraphics[width=0.95\textwidth]{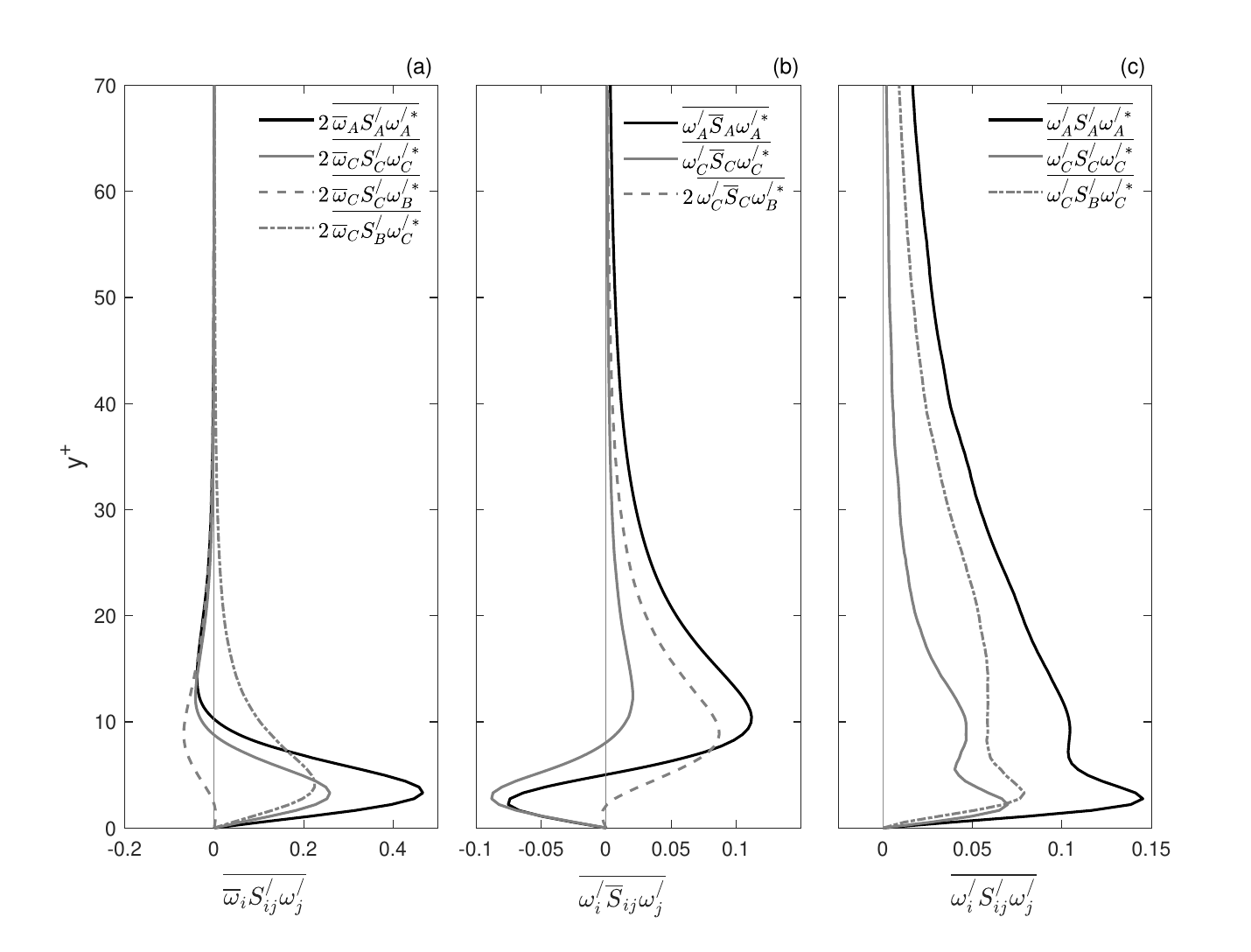}
 \caption{The most important term contributing to the budgets for $2 \overline{\overline{\omega}_{i}^{A}S_{ij}^{A\,'}\omega_{j}^{A\,'}}$ (a), $\overline{\omega_{i}^{A\,'}\overline{S}_{ij}^{A}\omega_{j}^{A\,'}}$ (b), and $\overline{\omega_{i}^{A\,'}S_{ij}^{A\,'}\omega_{j}^{A\,'}}$ (c) for $0 \le y^{+} \le 70$. For notational compactness, the $i$ and $j$ subscripts have been removed and the $B$ and $C$ superscripts changed to subscripts in the legends. The values for the terms formulated by \cite{motoori19} are given in black, the quantities only involving non-normal terms are shown as solid grey lines. Quantities that involve a mixture of normal and non-normal terms are shown as dotted or dash-dotted grey lines. All terms are normalized by $(\frac{1}{2}\langle ||\Omega_{A}||^{2} \rangle)^{\frac{3}{2}}$.}
 \label{fig.enstroprod}
  \end{figure}
  
Moving to index notation, we adopt a Reynolds decomposition where, for example, $\bm{\omega}^{A}=\overline{\bm{\omega}}^{A} + \bm{\omega}'^{\,A}$, with $\overline{\bm{\omega}}^{A}$ the mean vorticity vector at a constant distance from the wall, $y$, and $\bm{\omega}^{A \,'}$, the fluctuating part at the same height. Hence, given the spatial homogeneity in $x$ and $z$, averaging is undertaken in time, and in the longitudinal and transverse directions. It may then be shown \cite{motoori19} that the symmetries in a turbulent boundary-layer or channel flow lead to
\begin{align}
\notag{}
\overline{\omega}^{A}_{1} = \overline{\omega}^{A}_{2} &= 0\\
\overline{S}_{13}^{A} = \overline{S}_{31}^{A} = \overline{S}_{23}^{A}  = \overline{S}_{32}^{A}  = \overline{S}_{33}^{A} &= 0,
\label{eq.BL}
\end{align}
which results in an average enstrophy production given by 
\begin{equation}
\overline{\omega_{i}^{A}S_{ij}^{A}\omega_{j}^{A}} = 2 \overline{\overline{\omega}_{i}^{A}S_{ij}'^{\,A}\omega_{j}'^{\,A}} + \overline{\omega_{i}'^{\,A}\overline{S}_{ij}^{A}\omega_{j}'^{\,A}} + \overline{\omega_{i}'^{\,A}S_{ij}'^{\,A}\omega_{j}'^{\,A}}.
\label{eq.motoori}
\end{equation}
We refer to the last term on the right-hand side of (\ref{eq.motoori}) as the ``fluctuating enstrophy production''. The simulation by \cite{motoori19} employed a Reynolds number based on shear velocity and momentum thickness of 3170, and the patterns seen in their results (Fig. 5 of their paper) are replicated in our results as given by the black lines in Fig. \ref{fig.enstroprod}. In our figure, results are non-dimensionalized using the mean enstrophy from the wall to $y^{+} = 70$, the top of the domain considered in this study, i.e. $\frac{1}{2}\langle ||\Omega_{A}||^{2} \rangle^{\frac{3}{2}}$, where the angled braces are the average over all spatial directions including $y$. This value corresponded approximately to that observed on average at $y^{+} = 16$.

\cite{motoori19} showed that the first term on the right-hand side of (\ref{eq.motoori}) dominates the budget for $y^{+} \le 10$ and peaks at $y^{+} = 3.5$, with a small magnitude negative peak at $y^{+} = 14$. The second term on the right-hand side of (\ref{eq.motoori}) is typically the smallest magnitude term and is initially negative in sign until $y^{+} = 5$, reaching a negative peak at $y^{+} = 2.5$. It attains a positive peak at $y^{+} = 10$, which is about 50\% larger in magnitude than the negative peak. The fluctuating enstrophy production in panel (c) exhibits a positive peak at $y^{+} = 2.8$, and a secondary maximum at $y^{+} = 10$. This term then decays more slowly than the others and consequently, dominates the budget for $y^{+} > 30$. \cite{motoori19} showed that this is the case as a far out as at least $y^{+} = 1000$. However, in this study we focus on the more complex interplay for the various terms near the wall seen in the figure and, thus, focus attention on $y^{+} \le 70$.  

Applying the Schur decomposition to (\ref{eq.motoori}) results in the following sets of terms:
\begin{align}
\nonumber
\overline{\overline{\omega}_{i}^{A}S_{ij}'^{\,A}\omega_{j}'^{\,A}} &= \overline{\overline{\omega}_{i}^{B}S_{ij}'^{\,B}\omega_{j}'^{\,B}} + \overline{\overline{\omega}_{i}^{B}S_{ij}'^{\,B}\omega_{j}'^{\,C}} + \overline{\overline{\omega}_{i}^{B}S_{ij}'^{\,C}\omega_{j}'^{\,B}} + \overline{\overline{\omega}_{i}^{B}S_{ij}'^{\,C}\omega_{j}'^{\,C}}\\
&+ \overline{\overline{\omega}_{i}^{C}S_{ij}'^{\,B}\omega_{j}'^{\,B}} + \overline{\overline{\omega}_{i}^{C}S_{ij}'^{\,B}\omega_{j}'^{\,C}} + \overline{\overline{\omega}_{i}^{C}S_{ij}'^{\,C}\omega_{j}'^{\,B}} + \overline{\overline{\omega}_{i}^{C}S_{ij}'^{\,C}\omega_{j}'^{\,C}}
\label{eq.CVEP}
\end{align}

\begin{align}
\nonumber
\overline{\omega_{i}'^{\,A}\,\overline{S}_{ij}^{A}\omega_{j}'^{\,A}} &= 
\overline{\omega_{i}'^{\,B}\,\overline{S}_{ij}^{B}\omega_{j}'^{\,B}} + 2\,\overline{\omega_{i}'^{\,B}\,\overline{S}_{ij}^{B}\omega_{j}'^{\,C}} + \overline{\omega_{i}'^{\,B}\,\overline{S}_{ij}^{C}\omega_{j}'^{\,B}} \\
&+ 2\,\overline{\omega_{i}'^{\,C}\,\overline{S}_{ij}^{C}\omega_{j}'^{\,B}} + \overline{\omega_{i}'^{\,C}\,\overline{S}_{ij}^{B}\omega_{j}'^{\,C}} +
\overline{\omega_{i}'^{\,C}\,\overline{S}_{ij}^{C}\omega_{j}'^{\,C}}
\label{eq.ASEP}
\end{align}

\begin{align}
\nonumber
\overline{\omega_{i}'^{\,A}S_{ij}'^{\,A}\omega_{j}'^{\,A}} &= \overline{\omega_{i}'^{\,B}S_{ij}'^{\,B}\omega_{j}'^{\,B}} + 2\,\overline{\omega_{i}'^{\,B}S_{ij}'^{\,B}\omega_{j}'^{\,C}} + \overline{\omega_{i}'^{\,B}S_{ij}'^{\,C}\omega_{j}'^{\,B}}\\
&+ 2\,\overline{\omega_{i}'^{\,C}S_{ij}'^{\,C}\omega_{j}'^{\,B}} + \overline{\omega_{i}'^{\,C}S_{ij}'^{\,B}\omega_{j}'^{\,C}} + \overline{\omega_{i}'^{\,C}S_{ij}'^{\,C}\omega_{j}'^{\,C}}
\label{eq.FEP}
\end{align}

The great majority of the terms in these expansions do not have a significant impact on the enstrophy budget near the wall. Thus, in Fig. \ref{fig.enstroprod} we show each term from the left-hand side of (\ref{eq.CVEP})-(\ref{eq.FEP}) and the two or three terms from the right-hand side of these equations that contribute in practice to the near-wall dynamics. Inspection of Fig. \ref{fig.enstroprod}a shows that three terms of the eight on the right-hand side of (\ref{eq.CVEP}) are important: the purely non-normal term (grey, solid line), which mimics the collective behaviour of the terms on the left-hand side most closely, $2\,\overline{\overline{\omega}_{i}^{C}S_{ij}'^{\,B}\omega_{j}'^{\,C}}$ (grey, dot-dashed), which is positive throughout, and a negative contribution from $2\,\overline{\overline{\omega}_{i}^{C}S_{ij}'^{\,C}\omega_{j}'^{\,B}}$ (grey, dashed). The primary peak at $y^{+} = 4$ is due to a similar magnitude contribution from the two terms involving $\overline{\omega}_{i}^{C}$ and $\omega_{j}'^{\,C}$, while the negative peak at $y^{+} = 13$ is a consequence of $2\,\overline{\overline{\omega}_{i}^{C}S_{ij}'^{\,C}\omega_{j}'^{\,B}}$ and the change in sign of $2\,\overline{\overline{\omega}_{i}^{C}S_{ij}'^{\,C}\omega_{j}'^{\,C}}$ at $y^{+} = 8.5$. 

\begin{table}
\caption{Values for the constituents of the fluctuating enstrophy production, $\overline{\omega_{i}'^{\,A}S_{ij}'^{\,A}\omega_{j}'^{\,A}}$ shown in Fig. \ref{fig.enstroprod}c, and expressed as a percentage of the value for $\overline{\omega_{i}'^{\,A}S_{ij}'^{\,A}\omega_{j}'^{\,A}}$ at selected values for $y^{+}$.}
\centering
\begin{tabular}{ccccccc}
\noalign{\smallskip}
$y^{+}$ & $\overline{\omega_{i}'^{\,B}S_{ij}'^{\,B}\omega_{j}'^{\,B}}$ & $\overline{\omega_{i}'^{\,C}S_{ij}'^{\,C}\omega_{j}'^{\,C}}$ & $2\,\overline{\omega_{i}'^{\,B}S_{ij}'^{\,B}\omega_{j}'^{\,C}}$ & $\overline{\omega_{i}'^{\,B}S_{ij}'^{\,C}\omega_{j}'^{\,B}}$ & $2\,\overline{\omega_{i}'^{\,C}S_{ij}'^{\,C}\omega_{j}'^{\,B}}$  & $\overline{\omega_{i}'^{\,C}S_{ij}'^{\,B}\omega_{j}'^{\,C}}$ \\
\noalign{\smallskip}
3 & 0.6 & 43.7 & 1.8 & -1.0 & 0.6 & 54.3\\
5 & 2.3 & 37.0 & 4.5 & -3.0 & -0.6 & 59.8\\
10 & 7.1 & 44.5 & 5.6 & -7.7 & -5.5 & 56.0\\
30 & 14.3 & 17.2 & 10.9 & -7.8 & -2.2 & 67.6\\
70 & 18.6 & 11.1 & 22.2 & -3.6 & -2.6 & 54.3\\
\noalign{\smallskip}
\label{table.omASAomA}
\end{tabular}
\end{table}

Figure \ref{fig.enstroprod}b shows that just two of the six terms in (\ref{eq.ASEP}) are sufficient to explain the rather complex behaviour of this term. Apart from being slightly negative very near the wall, $2\,\overline{\omega_{i}'^{\,C}\,\overline{S}_{ij}^{\,C}\omega_{j}'^{\,B}}$ (grey, dashed line) is generally positive and drives the positive peak, while $2\,\overline{\omega_{i}'^{\,C}\,\overline{S}_{ij}^{\,C}\omega_{j}'^{\,C}}$ changes sign at $y^{+} = 8.5$ and drives the near-wall negative peak. Overall, the budget for (\ref{eq.FEP}) near the wall reduces to contributions from $\overline{\omega_{i}'^{\,C}S_{ij}'^{\,C}\omega_{j}'^{\,C}}$ (grey solid line) and $\overline{\omega_{i}'^{\,C}S_{ij}'^{\,B}\omega_{j}'^{\,C}}$ (grey dot-dashed line), with both of these terms positive throughout. The secondary maximum at $y^{+} = 12$ for $\overline{\omega_{i}'^{\,A}S_{ij}'^{\,A}\omega_{j}'^{\,A}}$ also seen in the results of  \cite{motoori19}) is much more pronounced for the purely non-normal term. The larger magnitude $\overline{\omega_{i}'^{\,C}S_{ij}'^{\,B}\omega_{j}'^{\,C}}$ term also exhibits two maxima, but these are both further from the wall, with the vertical offset between the maxima for the two terms acting to smooth the upper maximum for the overall fluctuating enstrophy production. It is also noticeable that the persistent positivity of $\overline{\omega_{i}'^{\,A}S_{ij}'^{\,A}\omega_{j}'^{\,A}}$ away from the wall is driven more by $\overline{\omega_{i}'^{\,C}S_{ij}'^{\,B}\omega_{j}'^{\,C}}$ than the purely non-normal term, which has decayed towards zero by $y^{+} = 70$. Thus, the driver for enstrophy production away from the wall is the local, fluctuating straining of the non-normal contribution to the fluctuating vorticity. This may also be seen in the last column of Table \ref{table.omASAomA} where this term accounts for over 50\% of the fluctuating enstrophy production at all heights, while the purely non-normal term decreases from 44\% near the wall to 11\% at $y^{+} = 70$. Consistent with this, further from the wall, the purely normal term (second column) increases dramatically in importance as one moves away from the wall, from a 0.6\% contribution at $y^{+} = 3$ to 18.6\% at $y^{+} = 70$

 \begin{figure}
 \centering
\includegraphics[width=0.95\textwidth]{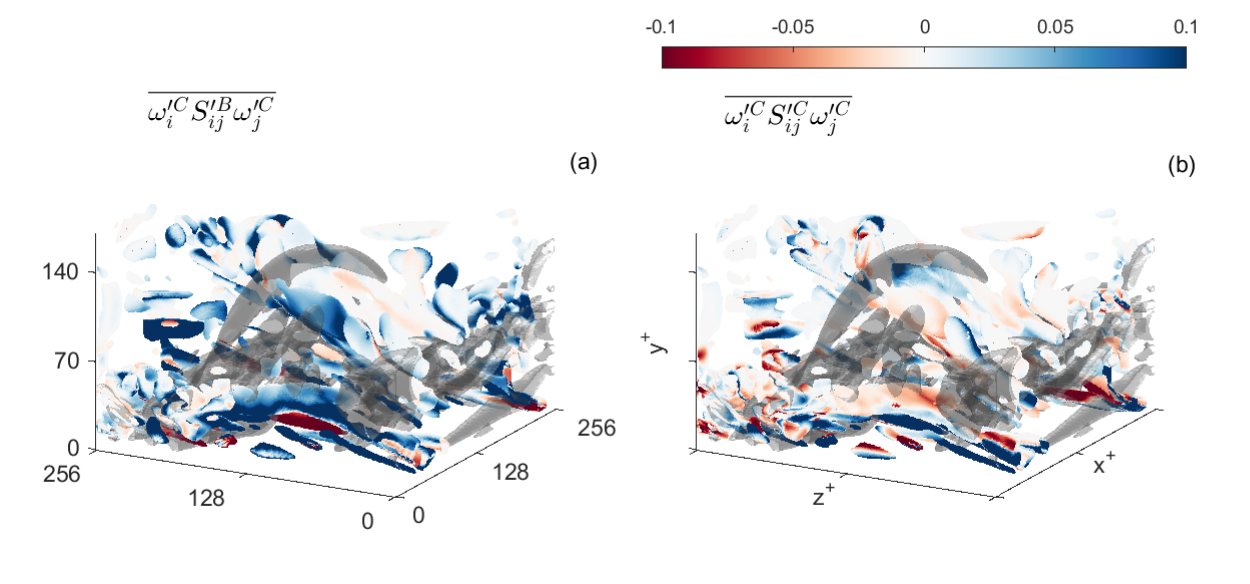}
 \caption{Snapshots of the same flow field shown in Fig. \ref{fig.QR}. The grey areas show locations with large values for $|Q|$. Otherwise, the colours reflect the values for the final two terms in (\ref{eq.FEP}) that are given as grey dash-dotted and grey lines, respectively in Fig. \ref{fig.enstroprod}c. These fields are draped onto regions of high $|R|$ and all terms are normalised by $(\frac{1}{2}\langle ||\Omega_{A}||^{2} \rangle)^{\frac{3}{2}}$.}
 \label{fig.eq45ef}
\end{figure}

An instantaneous snapshot of the two key terms in the fluctuating enstrophy production budget are shown in Fig. \ref{fig.eq45ef} superimposed onto regions of high $|R|$, with locations with high $|Q|$ shown in grey for reference. The general characteristics of the mean profiles may be discerned in the fields shown. For example, the general positivity of both terms (blue), the greater persistence of this positivity with height in the case of $\overline{\omega_{i}'^{\,C}S_{ij}'^{\,B}\omega_{j}'^{\,C}}$ in panel (a), and the greater predominance of negative production regions in panel (b), helping to explain the less positive mean profile for $\overline{\omega_{i}'^{\,C}S_{ij}'^{\,C}\omega_{j}'^{\,C}}$ above the viscous layer in Fig. \ref{fig.enstroprod}c.

That  Fig. \ref{fig.enstroprod}c and Table \ref{table.omASAomA} highlight the importance of $\overline{\omega_{i}'^{\,C}S_{ij}'^{\,B}\omega_{j}'^{\,C}}$ is interesting in terms of the analysis of HIT by \cite{k18}. Given there is no mean flow or preferred direction in such a flow, the positivity of $\overline{\omega_{i}'^{\,C}S_{ij}'^{\,B}\omega_{j}'^{\,C}}$ has a direct analogy to that study where this ``interaction production'' was largely responsible for driving the simultaneous positivity of the strain production and enstrophy production budgets in HIT. However, in a boundary-layer, given the small magnitude of the normal straining near the wall, it is not obvious that this term should dominate over the purely non-normal term particularly for small, $y^{+}$. In addition, the two maxima seen at $y^{+} = 3$ and $y^{+} = 10$ for $\overline{\omega_{i}'^{\,A}S_{ij}'^{\,A}\omega_{j}'^{\,A}}$ in Fig. \ref{fig.enstroprod}c are reflected in both the constituent terms shown in that panel, and the reasons for this are likely to be due to contributions from different parts of the vorticity and straining at different heights. Hence, in the next section we decompose the two most important terms in Fig. \ref{fig.enstroprod}c to determine which components of the strain rate and rotation rate tensors drive these terms. 

\section{The terms driving the fluctuating enstrophy production}

\subsection{Individual components of the fluctuating enstrophy production}

\begin{figure}
 \centering
\includegraphics[width=0.95\textwidth]{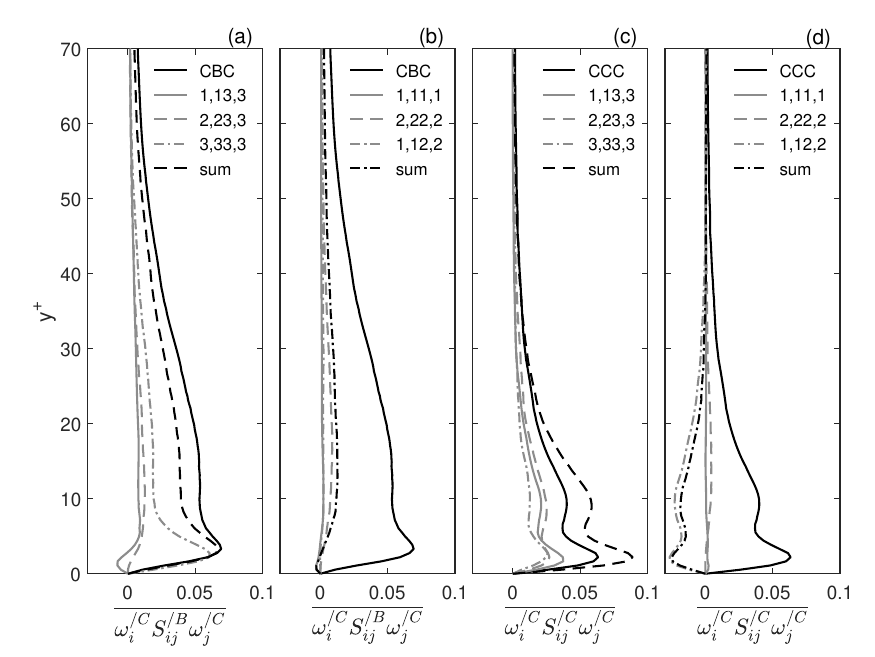}
 \caption{Vertical profiles for the constituent elements of the two dominant terms for the fluctuating enstrophy production. The components of $\overline{\omega_{i}^{'\,C}S_{ij}^{'\,B}\omega_{j}^{'\,C}}$ are given in (a) and (b) and $\overline{\omega_{i}^{'\,C}S_{ij}^{'\,C}\omega_{j}^{'\,C}}$ are shown in (c) and (d). In each case, the solid black line is equivalent to the appropriate case from Fig. \ref{fig.enstroprod}c and the other black line is the sum of the terms shown in grey. The legend indicates the nature of the component terms, which are shown in grey and, for example, ``1,13,3'' in panel (a) indicates $2\overline{\omega_{1}^{'\,C}S_{13}^{'\,B}\omega_{3}^{'\,C}}$. All terms are non-dimensionalized by $(\frac{1}{2}\langle ||\bm{\Omega}_{A}||^{2} \rangle)^{3/2}$.}
 \label{fig.OmCSBCOmC}
  \end{figure}
  
Figure \ref{fig.OmCSBCOmC}a,b show the profile of $\overline{\omega_{i}^{'\,C}S_{ij}^{'\,B}\omega_{j}^{'\,C}}$ from Fig. \ref{fig.enstroprod}c as a black line (CBC), with the six constituents of this term shown as grey lines, with three of the six terms in each panel. The sum of these three terms is then shown as a black dashed or dash-dotted line. This approach to the decomposition is repeated in Figure \ref{fig.OmCSBCOmC}c,d for 
$\overline{\omega_{i}^{'\,C}S_{ij}^{'\,C}\omega_{j}^{'\,C}}$. The components making up each line are given in the legend, where the commas delimit the vorticity from the straining and then from the vorticity and where the straining terms is an off-diagonal component it is implicit that the value has been multiplied by two due to the symmetry of the strain rate tensor. It is clear from panels (a) and (b) that the largest contribution to 
$\overline{\omega_{i}^{'\,C}S_{ij}^{'\,B}\omega_{j}^{'\,C}}$, particularly at the wall comes from $\overline{\omega_{3}^{'\,C}S_{33}^{'\,B}\omega_{3}^{'\,C}}$. The implication of this is that close to the wall the only really significant component of the fluctuating normal strain tensor is the transverse term. Panels (a) and (c) show the three terms that involve $\omega_{3}^{'\,C}$ and it is clear from (a) that their combined effect accounts for much of the relevant term from Fig. \ref{fig.enstroprod}c, with a deficit for $5 < y^{+} < 50$. 
Panel (b) shows that above the viscous sub-layer, the other three components all act in a positive manner to account for this deficit. In particular, the vertical term, $\overline{\omega_{2}^{'\,C}S_{22}^{'\,B}\omega_{2}^{'\,C}}$, is the principal term and the smooth increase in this term away from the wall explains the more diffuse peak to the second maximum for $\overline{\omega_{i}^{'\,C}S_{ij}^{'\,B}\omega_{j}^{'\,C}}$ compared to $\overline{\omega_{i}^{'\,C}S_{ij}^{'\,C}\omega_{j}^{'\,C}}$ in Fig. \ref{fig.enstroprod}c.

The primary difference between panel (c) and panel (a) is that the sum of the three terms shown exceeds the values for the term itself, implying negative contributions on average from the remaining three terms, which are shown in panel (d). Hence, the unexpected dominance of $\overline{\omega_{i}^{'\,C}S_{ij}^{'\,B}\omega_{j}^{'\,C}}$ given how small the normal parts of the tensor are near the wall is because of negative contributions to the budget for the term that dominates panel (d), $\overline{\omega_{1}^{'\,C}S_{12}^{'\,C}\omega_{2}^{'\,C}}$. Hence, it is the non-normal contribution to the fluctuating shearing in the dominant longitudinal-vertical plane that acts to reduce the purely non-normal contribution to fluctuating enstrophy production. The three terms in panel (c) are of similar magnitude, in contrast to panel (a). Of these,  $\overline{\omega_{1}^{'\,C}S_{13}^{'\,C}\omega_{3}^{'\,C}}$ is the largest very close to the wall, with $\overline{\omega_{2}^{'\,C}S_{23}^{'\,C}\omega_{3}^{'\,C}}$ greater for $y^{+} > 5$. Thus, the important positive and negative contributions to $\overline{\omega_{i}^{'\,C}S_{ij}^{'\,C}\omega_{j}^{'\,C}}$ all involve off-diagonal components of the straining, in marked contrast to the normal tensor in panels (a) and (b).

\begin{figure}
 \centering
\includegraphics[width=0.95\textwidth]{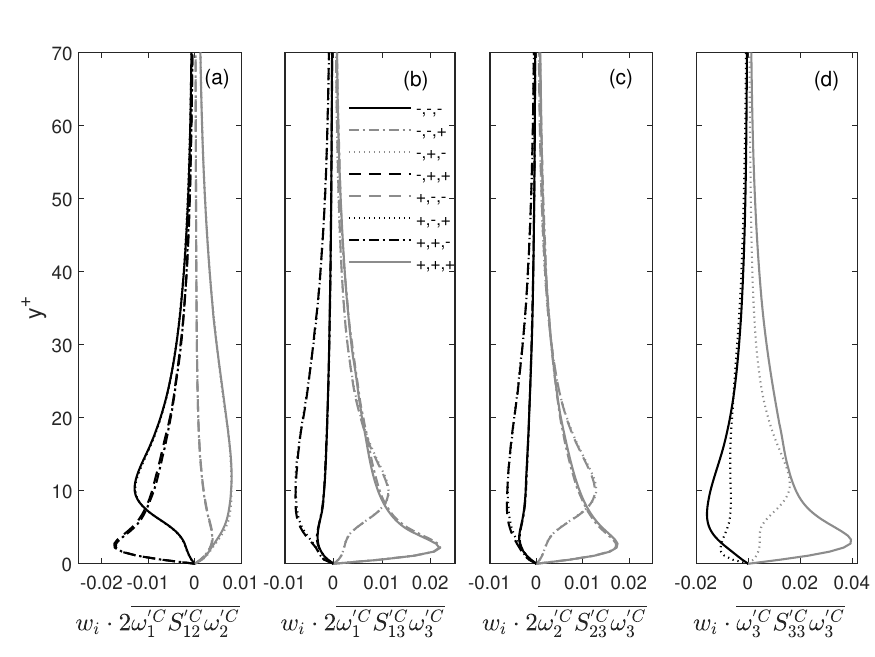}
 \caption{Vertical profiles for four components of  $\overline{\omega_{i}^{'\,C}S_{ij}^{'\,C}\omega_{j}^{'\,C}}$ are given in each panel. Each contains separate profiles conditioned on the signs of the respective vorticity and straining terms where, for example, the legend item ``-,+,+'' (black, dashed line) indicates that $\omega_{i}^{'\,C} < 0$, $S_{ij}^{'\,C} > 0$ and $\omega_{j}^{'\,C} > 0$. Each profile is weighted by the relative frequency of occurrence of the particular sign combinations, $w_{i}$ at that $y^{+}$ such that the magnitude of the contribution is proportional to that in Fig. \ref{fig.OmCSBCOmC}. All terms are non-dimensionalized by $(\frac{1}{2}\langle ||\bm{\Omega}_{A}||^{2} \rangle)^{3/2}$.
}
 \label{fig.vSv_signs}
  \end{figure}

To gain greater insight into the four terms that drive the non-normal enstrophy production, 
$\overline{\omega_{i}^{'\,C}S_{ij}^{'\,C}\omega_{j}^{'\,C}}$ in Fig. \ref{fig.OmCSBCOmC}c,d we condition each of these on the signs of fluctuating vorticity and strain rate, and examine the contribution from each in Fig. 
 \ref{fig.vSv_signs}. Hence, their are eight lines shown in panels (a)-(c) and four lines in (d), where we consider $\overline{\omega_{3}^{'\,C}S_{33}^{'\,C}\omega_{3}^{'\,C}}$. The values plotted are the mean values for each terms at a given $y^{+}$ multiplied by $w_{i}$, the relative frequency of one of the eight (or four) states. This means that the lines shown are proportional to their contribution to Fig. \ref{fig.OmCSBCOmC}. It is clear that in  Fig. 
 \ref{fig.vSv_signs}a-c there are four pairs of values and given that the sign of the contribution is given by the product of the signs of the three terms, after this the nature of the profile is then given by sign of $S_{12}^{'\,C}$ in (a) and $\omega_{3}^{'\,C}$ in (b,c), while the sign of $S_{33}^{'\,C}$ drives the nature of the profile in (d). The results in (d) are approximately double in magnitude those in the other panels despite the prefactor of a 2 in the other cases, indicating the important role of the transverse direction for non-normal enstrophy production near the wall. The positive contributions in (b)-(d) and negative contributions in (a) all consist of a profile that peaks in the viscous sub-layer and another that peaks at $y^{+} \sim 10$ explaining the double-peaked profile for the fluctuating enstrophy production overall \cite{motoori19}, as well as the component terms in Fig. \ref{fig.enstroprod}c. The profiles with the viscous sub-layer peak in (b)-(d) are associated with $\omega_{3}^{'\,C} > 0$ and those with the $y^{+} \sim 10$ peak with $\omega_{3}^{'\,C} < 0$, highlighting a fundamental change in the nature of the rotational component of the flow with a transverse axis. 

\subsection{Vorticity, straining and their mutual correlation}
 \begin{figure}
 \centering
\includegraphics[width=0.95\textwidth]{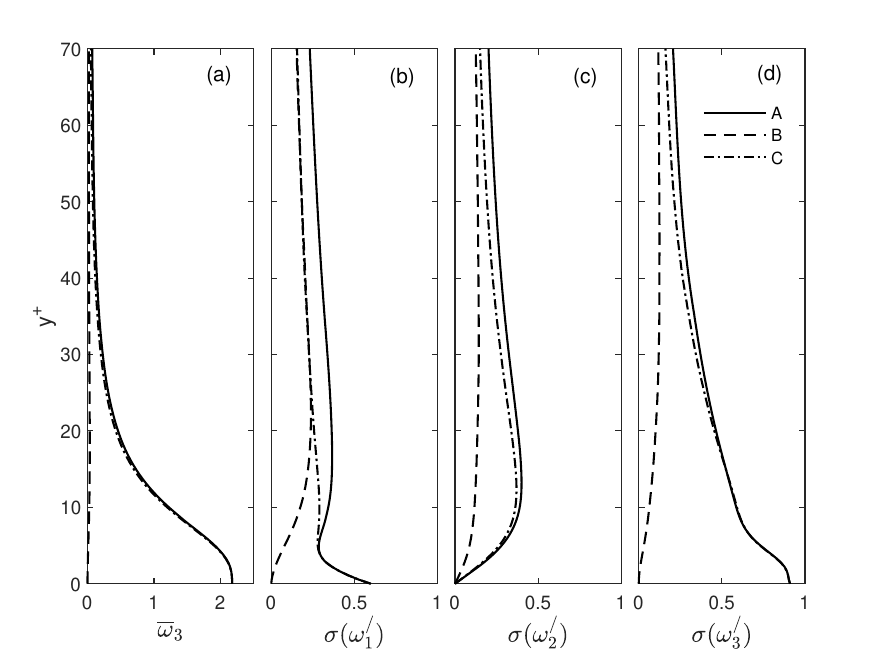}
 \caption{Vertical profiles for the individual components of the vorticity and their Schur-based variants (different lines), with the non-zero mean component shown in (a) and the standard deviation of the fluctuating components given in panels (b) to (d). All terms are normalized by $(\frac{1}{2}\langle ||\bm{\Omega}_{A}||^{2} \rangle)^{\frac{1}{2}}$.}
 \label{fig.vortprof}
  \end{figure}
 
The contributions of different terms to the mean and fluctuating vorticities are shown in Fig. \ref{fig.vortprof}, with the standard deviation, $\sigma(\ldots)$, used to summarize the fluctuating values. It is clear that the normal contributions are negligible at the wall for all components but, in the case of $\sigma(\omega_{1}^{'\,B})$ this grows to be as important as $\sigma(\omega_{1}^{'\,C})$ by $y^{+} \sim 20$. While the profile for $\sigma(\omega_{1}^{'\,C})$ exhibits an initial sharp decay with height before a recovery at $y^{+} = 5$ and then a slow increase, that for $\sigma(\omega_{2}^{'\,C})$ is zero at the wall and then increases more rapidly. This explains why in Fig.\ref{fig.OmCSBCOmC}c values for $\overline{\omega_{1}^{'\,C}S_{13}^{'\,C}\omega_{3}^{'\,C}}$ are greater at the wall and those for $\overline{\omega_{2}^{'\,C}S_{23}^{'\,C}\omega_{3}^{'\,C}}$ are greater for $y^{+} > 5$. This effect can also be seen from the different magnitude of the positive contributions in panels (b) and (c) of Fig. \ref{fig.vSv_signs}. We have already noted a change in the sign of $\omega_{3}^{'\,C}$ for the features that dominate the positive non-normal enstrophy production and this is implied by the inflection in the profile for $\sigma(\omega_{3}^{'\,C})$ in Fig. \ref{fig.vortprof}d at $y^{+} \sim 7$. 

\begin{figure}
 \centering
\includegraphics[width=0.95\textwidth]{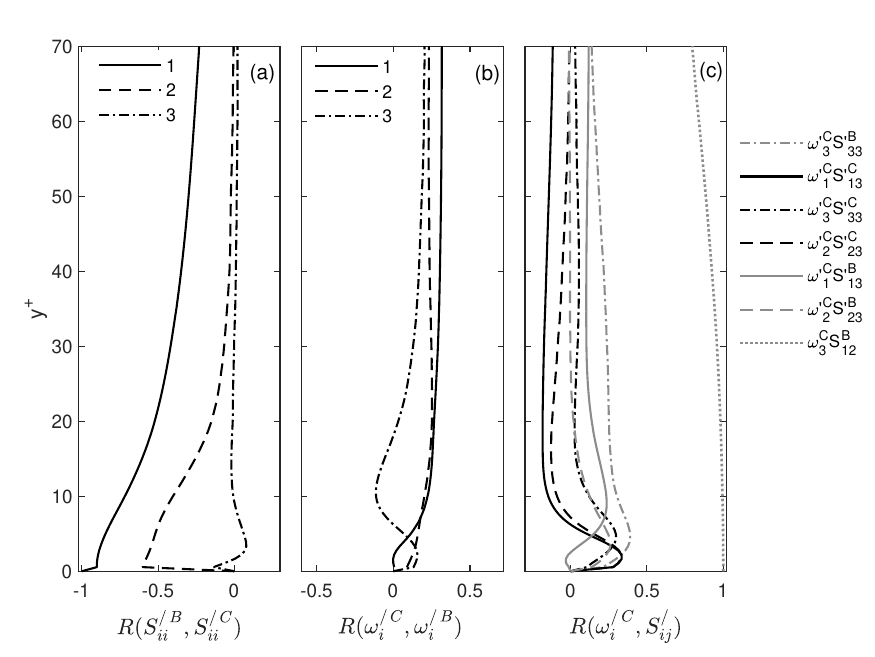}
 \caption{Vertical profiles for the linear correlation between selected components of the straining and vorticity for the normal and non-normal tensors.
}
 \label{fig.sv_corr}
  \end{figure}

For $y^{+} \lesssim 30$ the fluctuating non-normal vorticity is dominated by $\sigma(\omega_{3}^{'\,C})$, which explains why all the net positive contributions in Fig. \ref{fig.vSv_signs} involve this component of the vorticity. Aspects of the enstrophy production as well as the relation between the normal and non-normal tensors can be seen in Fig. \ref{fig.sv_corr}. Given the earlier discussion about the change in nature of $\omega_{3}^{'\,C}$, there is a change in sign of the correlation, $R(\omega_{3}^{'\,B},\omega_{3}^{'\,C})$, in panel (b), which is weakly positive near the wall and negative at $y^{+} = 10$. In contrast, $R(\omega_{2}^{'\,B},\omega_{2}^{'\,C})$ is weakly positive throughout the near-wall region and $R(\omega_{1}^{'\,B},\omega_{1}^{'\,C}) = 0$ in the very near wall region, below the inflection for the $\sigma(\omega_{1}^{'\,C})$ profile in Fig. \ref{fig.vortprof}b and then increases dramatically over $5 < y^{+} < 10$ as the normal component grows in magnitude. This transition in the behaviour of $\omega_{3}^{'\,C}$ is also seen in the change in signs of $R(\omega_{1}^{'\,C},S_{12}^{'\,C})$ (black, solid line) and $R(\omega_{2}^{'\,C},S_{23}^{'\,C})$ (black, dashed line) in Fig. \ref{fig.sv_corr}c at $y^{+} \sim 7$. The correlation for $R(\omega_{3}^{'\,C},S_{33}^{'\,C})$ in contrast remains positive, but is weaker in magnitude. Hence, the large contribution from this term seen in Fig. \ref{fig.vSv_signs}d are driven by the large values for the respective terms rather than a strong alignment. 

Given earlier results demonstrating that $\overline{\omega_{i}^{'\,C}S_{ij}^{'\,B}\omega_{j}^{'\,C}}$ is typically greater in magnitude near the wall than $\overline{\omega_{i}^{'\,C}S_{ij}^{'\,C}\omega_{j}^{'\,C}}$, and the specific contribution from $\overline{\omega_{3}^{'\,C}S_{33}^{'\,C}\omega_{3}^{'\,C}}$ in Fig. \ref{fig.OmCSBCOmC}a, it is not surprising that the largest correlation of terms that contribute to the enstrophy production is $R(\omega_{3}^{'\,C},S_{33}^{'\,B})$ (grey, dot-dashed line) in Fig. \ref{fig.sv_corr}c. 
However, we also show $R(\omega_{3}^{'\,C},S_{12}^{'\,C})$ (grey, dotted line) is close to 1 for $y^{+} < 20$ as this explains why in Fig. \ref{fig.vSv_signs}a the nature of the profiles is controlled by the sign of $S_{12}^{'\,C}$, while $\omega_{3}^{'\,C}$ fulfils a similar role in Fig. \ref{fig.vSv_signs}b,c. Given the dominance of $\omega_{3}^{'\,C}$ at the wall, which is linked directly to the dominant plane of shearing, then that these two terms are highly correlated until vorticity develops in other components of the non-normal tensor, or a significant normal component to the dynamics develops, means that the same mechanisms drive both the positive and negative contributions to the non-normal enstrophy production budget in panels (c) and (d) of Fig. \ref{fig.OmCSBCOmC}. 

The final key feature of Fig. \ref{fig.sv_corr} is the correlation between the diagonal components of the strain rate for the normal and non-normal tensors. While any correlation between $S_{33}^{'\,B}$ and $S_{33}^{'\,C}$ is negligible, both $R(S_{11}^{'\,B},S_{11}^{'\,C})$ and $R(S_{22}^{'\,B},S_{22}^{'\,C})$ are highly significant. For the former this is the case throughout the near-wall region while for the latter correlation decays to close to zero by the inertial regime at $y^{+} = 30$. This very strong anti-correlation between $S_{11}^{'\,B}$ and $S_{11}^{'\,C}$ provides a mechanism for the non-local effects of shearing and surrounding vortices, which impact on the non-normal part of the tensor, to influence the local dynamics. An induced effect on $S_{11}^{'\,B}$ will then have ramifications for the other components of the local dynamics due to incompressibility.

\begin{figure}
 \centering
\includegraphics[width=0.95\textwidth]{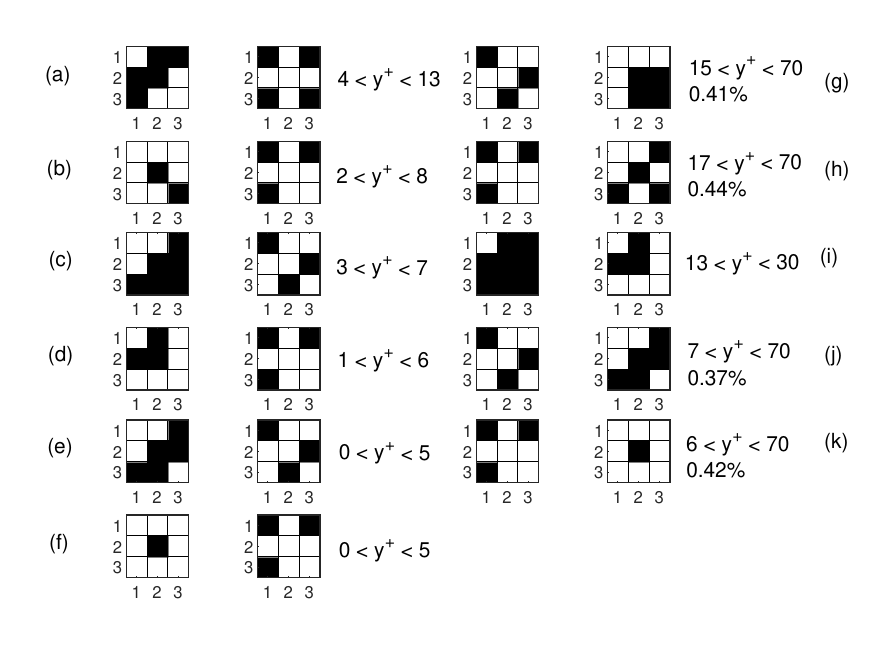}
 \caption{The most common strain rate tensor configurations as a function of $y^{+}$. For each pair of tensors, the left-hand panel of each set shows $\bm{S}_{B}^{/}$ and the right is $\bm{S}_{C}^{/}$. Positive values for the strain rates are shown in white and negative are in black. Four cases arise for a significant proportion of the range of heights considered and their average percentage occurrence for $50 < y^{+} < 70$ is stated for these cases.  
 }
 \label{fig.SBC}
  \end{figure}

Figure \ref{fig.SBC} indicates the most common fluctuating strain tensor sign configurations for $\bm{S}_{B}^{/}$ and $\bm{S}_{C}^{/}$ as a function of $y^{+}$ with positive values in white and negative in black. Panels (a) to (f) focus on the dominant cases for $y^{+} < 13$ while (g) to (k) are cases that are present near the wall, but typically come to dominate in the inertial regime. For $y^{+} > 50$ the same rank order of the four most common states arose and the average percentage of occurrences for these cases is given in panels (g), (h), (j) and (k). These percentages may seem small, but given the number of permutations of the signs for these two matrices, the values quoted are about twelve times the anticipated neutral percentage of $100(3 \times 2^{10})^{-1}\,\%$. The most obvious result is that for (a)-(f) we have $S_{11}^{/\,B} > 0$ and $S_{11}^{/\,C} < 0$, with the opposite the case for the vertical straining, while the four cases in the right-hand column that extend in influence as high as $y^{+} = 70$ have the opposite configuration. Hence, the results are more subtle than a simple anti-correlation as inferred from Fig. \ref{fig.sv_corr}a as the nature of this changes with height. In fact, the height at which the cases in (g) and (h) first emerge as important ($y^{+} \sim 16$) corresponds to the inflection point in the correlations for $S_{11}^{/\,B}$ and $S_{11}^{/\,C}$ and $S_{22}^{/\,B}$ and $S_{22}^{/\,C}$ in Fig. \ref{fig.sv_corr}a. In addition to these components of the straining changing in sign with height, dominant patterns for the straining shift between $\bm{S}_{B}^{/}$ and $\bm{S}_{C}^{/}$. Hence, panels (b), (d) and (f) have $\bm{S}_{C}^{/}$ with positive values with the exception of $S_{11}^{/\,C}$ and $S_{13}^{/\,C}$ (or $S_{31}^{/\,C}$). This then becomes a characteristic pattern for $\bm{S}_{B}^{/}$ further from the wall as seen in (h) and (k); panels (c) and (e) have $\bm{S}_{C}^{/}$ with positive values with the exception of $S_{11}^{/\,C}$ and $S_{23}^{/\,C}$ (or $S_{32}^{/\,C}$), which is then the case for (g) and (j). Given these results, the most unusual case is in panel (i) which is the only one where $S_{11}^{/\,B} > 0$ for the cases further from the wall and the only one where $S_{11}^{/\,B}$ and $S_{11}^{/\,C}$ have the same sign. Indeed in this case, all terms involving the longitudinal and vertical terms have the same sign for the normal and non-normal tensors and this is the only case shown where $S_{12}^{/\,C} < 0$. Thus, events with a strong negative fluctuating shear in the dominant plane arise in the transitional region and these are sufficiently large fluctuations in $\bm{S}_{A}^{/}$ that both the normal and non-normal tensors have the same sign. Negative values for all normal terms but $S_{11}^{/\,B}$ imply the formation of longitudinal, tube-like structures in this case.

The results in Fig. \ref{fig.vSv_signs}d show that the largest single contribution to the fluctuating, non-normal enstrophy production arises near the wall where both $\omega_{3}^{/\,C}$ and $S_{33}^{/\,C}$ are positive. This may be inferred from Fig. \ref{fig.SBC} as panels (b) to (f) (the cases restricted to $y^{+} < 8$) all indicate that $S_{33}^{/\,C} > 0$, while Fig. \ref{fig.sv_corr}c shows that in this region, the correlation between $\omega_{3}^{/\,C}$ and $S_{12}^{/\,B}$ is close to +1, and $S_{12}^{/\,B} > 0$ arises in (b) and (d)-(f). Panel (a) shows that the arrangement that peaks for $4 < y^{+} < 13$ is that with $S_{33}^{/\,C} < 0$ and $S_{12}^{/\,B} < 0$ and this corresponds to the negative peak in non-normal fluctuating enstrophy production in Fig. \ref{fig.sv_corr}c. The other positive peak in Fig. \ref{fig.sv_corr}c extends from $5 < y^{+} < 30$ with a maximum at $y^{+} \sim 12$. This requires $\omega_{3}^{/\,C} < 0$ and $S_{33}^{/\,C} > 0$, and using the correlation between $\omega_{3}^{/\,C}$ and $S_{12}^{/\,B}$ once more, we find that this arises for the configuration in panel (i), which was independently identified as being important for $13 < y^{+} < 30$ and was shown to correspond to strong negative fluctuating shear events in the previous paragraph. 

\subsection{Large fluctuating shear events}

\begin{figure}
 \centering
\includegraphics[width=0.95\textwidth]{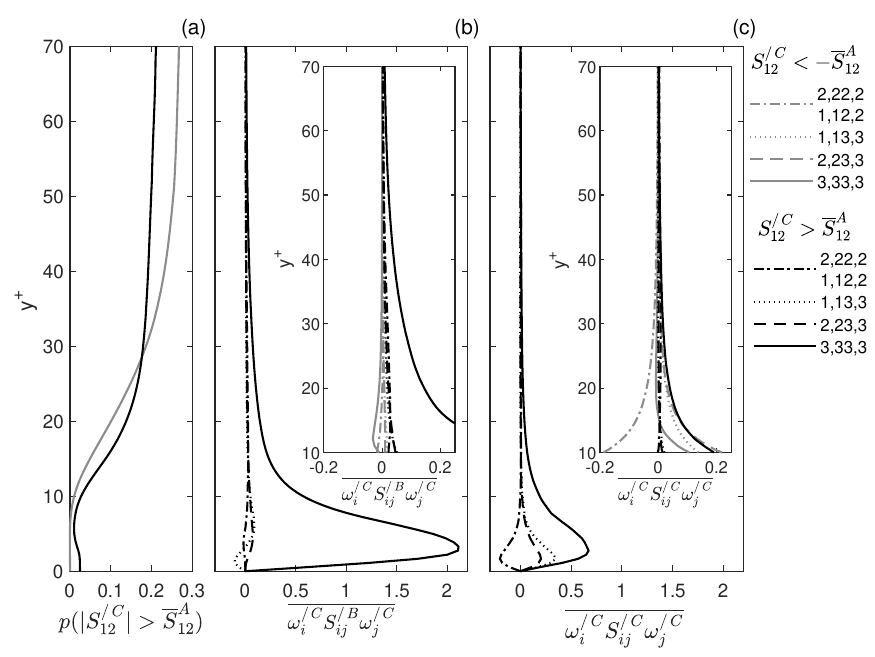}
 \caption{Vertical profiles of key components of the two dominant contributions to the fluctuating enstrophy production where $\vert S_{12}^{/\,C}\vert > \overline{S}_{12}^{/\,A}$. Panel (a) shows the probability of exceeding this threshold while (b) and (c) show the fluctuating enstrophy production for terms involving $S_{ij}^{/\,B}$ and $S_{ij}^{/\,C}$, respectively. Throughout, negative occurrences for the threshold exceedance are in grey and positive are in black. The insets in each panel focus on $y^{+} \ge 10$ where the number of negative threshold exceedances becomes significant (hence, the main panels only show positive exceedances). 
 Values are non-dimensionalized by $\frac{1}{2}\langle \Vert\bm{\Omega}_{A}\Vert^{2} \rangle^{3/2}$.}
\label{fig.S12Ccond}
  \end{figure} 

The largest magnitude fluctuating vorticity term is shown in Fig.\ref{fig.vortprof}c to be $\omega_{3}^{/\,A}$, which for most of the near-wall region is dominated by $\omega_{3}^{/\,C}$, which implies $S_{12}^{/\,C}$ is the largest term in the fluctuating strain rate tensors (which is the case, although this is not shown above). Given the dominant contribution to the mean dynamics of $\overline{S}_{12}^{A}$, we consider large fluctuating events to be those where $\vert S_{12}^{/\,C}\vert > \overline{S}_{12}^{/\,A}$ and then consider the average fluctuating enstrophy production values for the negative and positive exceedances in Fig. \ref{fig.S12Ccond}. Such extreme events will drive the mean statistics determined over all points, as seen earlier.    

Panel (a) shows the empirical probability of exceeding the threshold as a function of $y^{+}$. Negative occurrences essentially do not arise until $y^{+} > 10$. They then increase rapidly such that, at the start of the inertial regime at $y^{+} \sim 30$ they are as likely as the positive occurrences. Further into the flow, as the mean shearing declines, the negative occurrences become more probable, reflecting the effectiveness of ejection events in a boundary-layer \cite{nezu77}. The fluctuating enstrophy production that results from the positive threshold exceedances is shown in panels (b) and (c) with four terms considered in each case (those that dominate in Fig. \ref{fig.OmCSBCOmC}). Three of these are similar in terms of their components for (b) and (c) with the difference being the contribution from the normal or non-normal tensor. One term, shown by dot-dashed lines, does differ however and as the legend indicates, this is $\overline{\omega_{2}^{/\,C}S_{22}^{/\,B}\omega_{2}^{/\,C}}$ in (b) and $\overline{\omega_{1}^{/\,C}S_{12}^{/\,C}\omega_{2}^{/\,C}}$ in (c). In both panels (b) and (c) the insets highlight the behaviour for $y^{+} > 10$ and includes the negative threshold exceedances (in grey), which did not really arise nearer the wall as seen in panel (a). The near wall dominance of $\overline{\omega_{3}^{/\,C}S_{33}^{/\,B}\omega_{3}^{/\,C}}$ for positive threshold exceedances (solid, black line in panel (b)) is even clearer here than in earlier results, while the negative contribution to the purely non-normal term from $\overline{\omega_{1}^{/\,C}S_{12}^{/\,C}\omega_{2}^{/\,C}}$, seen in Fig. \ref{fig.vSv_signs}a is shown in panel (c) to be a stronger negative enstrophy production contribution than any other term in panels (b) or (c) as expected from that earlier result. The inset to panel (c) shows that for $y^{+} > 10$ there is only one significant negative contribution to the purely non-normal term and that this is $\overline{\omega_{1}^{/\,C}S_{12}^{/\,C}\omega_{2}^{/\,C}}$ for the negative threshold exceedance. However, an important distinction between panels (b) and (c) is that in the former, the negative threshold exceedances result in negative enstrophy production, while in (c) only the term mentioned is negative with, to a good approximation $\overline{\omega_{1}^{/\,C}S_{12}^{/\,C}\omega_{2}^{/\,C}} = -\overline{\omega_{2}^{/\,C}S_{23}^{/\,C}\omega_{3}^{/\,C}}$ for the $S_{12}^{/\,C} < -\overline{S}_{12}^{A}$ threshold. Thus, in the near-wall region, a strong non-normal contribution to the fluctuating shear results in general in positive enstrophy production and, hence, vortex development. The one term that consistently acts to reduce fluctuating enstrophy production (irrespective of the sign of the fluctuating shear) is $\overline{\omega_{1}^{/\,C}S_{12}^{/\,C}\omega_{2}^{/\,C}}$. Hence, this is associated with vortex decay and dissipation. The sign-by-sign breakdown in Fig. \ref{fig.vSv_signs}a shows that the primary negative contributions to this term can arise for either sign for 
$S_{12}^{/\,C}$ with the positive occurrences nearer the wall, and this is borne out by the conditioning for the extremal cases in Fig. \ref{fig.S12Ccond}c. Thus, the direct impact of large fluctuations in the shearing is dissipative in nature, but on average in the near wall region, associated indirect interactions lead to positive fluctuating enstrophy production when the fluctuations are large.

\begin{figure}
 \centering
\includegraphics[width=0.95\textwidth]{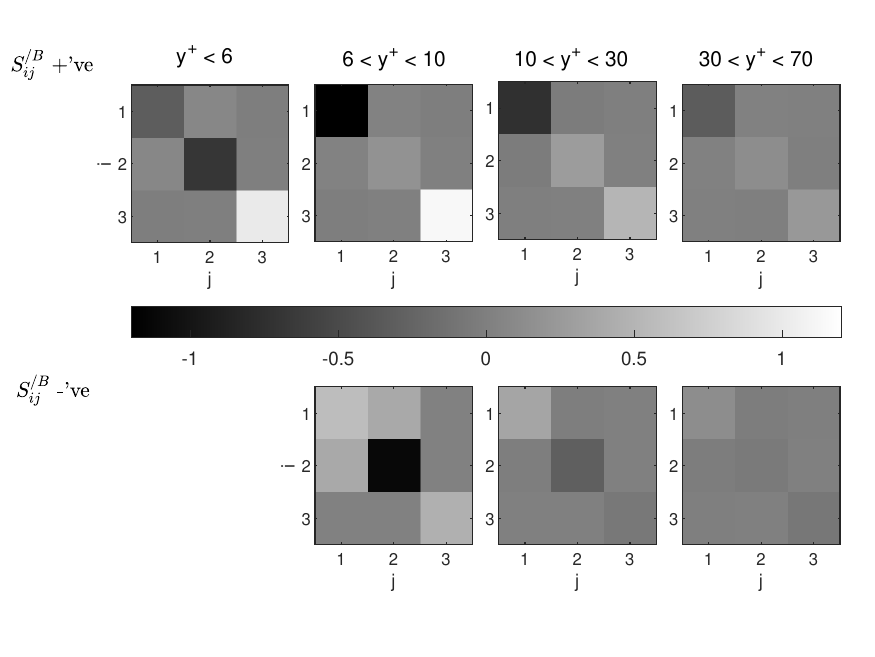}
 \caption{Typical $S_{ij}^{/\,B}$ tensors at four different $y^{+}$ ranges for the cases where $\vert S_{12}^{/\,C}\vert > \overline{S}_{12}^{/\,A}$. The lower row are the tensors for $S_{12}^{/\,C} < -\overline{S}_{12}^{/\,A}$, while the upper row is for  $S_{12}^{/\,C} > \overline{S}_{12}^{/\,A}$.  Values are normalized by $S_{33}^{/\,B}$ for $y^{+} < 6$.}
\label{fig.SBextrem}
  \end{figure} 

Given the threshold is defined in terms of $\vert S_{12}^{/\,C}\vert$, the fluctuating non-normal strain rate tensors that exceed the threshold are dominated by $\vert S_{12}^{/\,C}\vert$, with negative and positive values for negative and positive exceedances of the threshold, respectively. However, the nature of $S_{ij}^{/\,B}$ is not set by the nature of the threshold and Fig. \ref{fig.SBextrem} shows these tensors for negative (bottom row) and positive (top row) exceedances of this threshold and for four ranges of $y^{+}$ values chosen on the basis of the results in Fig. \ref{fig.S12Ccond}. No results are given for $y^{+} < 6$ for the negative threshold because, as shown in Fig. \ref{fig.S12Ccond}a, such occurrences essentially do not arise. Because of the importance of $\overline{\omega_{3}^{'\,C}S_{33}^{'\,B}\omega_{3}^{'\,C}}$ for the dynamics, the values in Fig. \ref{fig.SBextrem} are normalized with respect to $S_{33}^{'\,B}$ for $y^{+} < 6$. 

For positive threshold exceedances, $S_{33}^{'\,B}$ is the largest magnitude term for $y^{+} < 6$. However, above this, the negative values for $S_{11}^{'\,B}$ are greater in magnitude, although the limited correlation between $S_{11}^{'\,B}$ and $\omega_{1}^{'\,C}$ results in very little enstrophy production from this strain component (insufficient to feature in Fig. \ref{fig.S12Ccond}). It is the diagonal components that dominate the normal strain tensor for positive threshold exceedances in marked contrast to $\vert S_{12}^{/\,C}\vert$ dominating the non-normal tensor (for $30 < y^{+} < 70$ this component is twelve times greater than any other; for $y^{+} < 6$ it is thirty times greater). Thus, even in the presence of extreme fluctuating shearing, the normal tensor is promoting axial straining, the nature of which varies with height. In the top row of Fig. \ref{fig.SBextrem} we see that for $y^{+} < 6$ there is extensional behaviour in the transverse direction, with the compression in the other two directions driven by the vertical component. By $6 < y^{+} < 10$ $S_{11}^{'\,B}$ is of a similar magnitude and we have close to two-dimensional behaviour. For $y^{+} > 10$ as $S_{33}^{'\,B}$ declines in magnitude relative to $S_{11}^{'\,B}$ we move towards the case where foreshortening in the longitudinal direction induces extension in the transverse direction that is about twice the extension in the vertical. For the negative exceedances we have a very different dynamics where, for $y^{+} < 10$ vertical compression result in approximately axisymmetric extension in the longitudinal/transverse plane. For $y^{+} > 10$ we have a change in sign for $S_{33}^{'\,B}$ with this term initially weakly negative such that for $10 < y^{+} < 30$ we have approximately two-dimensional behaviour (but in a different plane to the result for $6 < y^{+} < 10$ for the positive exceedances). For $30 < y^{+} < 70$ we have approximately axisymmetric behaviour with tube-like structures that are longitudinally oriented.

\section{Conclusion}
Following a Reynolds decomposition of the enstrophy production equation, \cite{motoori19} found that the fluctuating enstrophy production term was the most important to the budget for enstrophy production in the near-wall region, but the profile they obtained (see Fig. \ref{fig.enstroprod}c) has a complex shape, with a double peak, implying different terms are contributing to the fluctuating enstrophy production at different heights. To investigate this phenomenon in greater detail, we have undertaken a Schur decomposition of the velocity gradient tensor to isolate the normal (denoted by a ``B'') and non-normal (denoted by a ``C'') contributions to the strain rate and rotation rate tensors \cite{k18,xu19}. Using this approach we have found that four particular terms dominate: $\overline{\omega_{3}^{'\,C}S_{33}^{'\,B}\omega_{3}^{'\,C}}$, $\overline{\omega_{1}^{'\,C}S_{13}^{'\,C}\omega_{3}^{'\,C}}$, $\overline{\omega_{2}^{'\,C}S_{23}^{'\,C}\omega_{3}^{'\,C}}$ and $\overline{\omega_{3}^{'\,C}S_{33}^{'\,C}\omega_{3}^{'\,C}}$, with the former the most important. That this term is greater than the latter is particularly surprising given the strong mean and fluctuating shear near the wall induces a non-local response on individual flow structures that is captured by the non-normal terms. However, not only are $\omega_{3}^{'\,C}$ and $S_{33}^{'\,B}$ strongly correlated (Fig. \ref{fig.sv_corr}c)  but one component of the purely non-normal contribution to the fluctuating enstrophy production, $\overline{\omega_{1}^{'\,C}S_{12}^{'\,C}\omega_{2}^{'\,C}}$ is negative on average, reducing the positive contribution from $\overline{\omega_{i}^{'\,C}S_{ij}^{'\,C}\omega_{j}^{'\,C}}$ relative to $\overline{\omega_{i}^{'\,C}S_{ij}^{'\,B}\omega_{j}^{'\,C}}$.

To explain the double-peaked nature of the profiles we investigated the contributions of the above terms based on the signs of the vorticity and strain components. Figure \ref{fig.vSv_signs} shows that for the purely non-normal fluctuating enstrophy production, the peak in, or close to the viscous sub-layer is driven by cases where either all terms have a positive sign or where $S_{i3}^{'\,C} < 0$ but $\omega_{3}^{'\,C} > 0$. The peak closer to $y^{+} = 10$ arises where $\omega_{3}^{'\,C} < 0$. In contrast, Fig. \ref{fig.OmCSBCOmC}a,b show that $\overline{\omega_{3}^{'\,C}S_{33}^{'\,B}\omega_{3}^{'\,C}}$ dominates near the wall and is still significant at greater heights, but the increase in magnitude of other components of the fluctuating normal strain tensor drives a more diffuse upper peak for $10 < y^{+} < 20$. 

Further investigation of the straining aspects of the enstrophy production show that the most common tensor arrangements for $y^{+} \lesssim  10$ have $S_{11}^{'\,B} > 0$ and $S_{11}^{'\,C} < 0$ and this swaps for $y^{+} \gtrsim  10$ (Fig. \ref{fig.SBC}). However, when we focus on the extreme cases, defined as where   
$\vert S_{12}^{'\,C}\vert > \overline{S}_{12}^{A}$, i.e. fluctuating non-normal shear is greater than mean shear, which will be dominating the mean values, we find marked differences depending on $y^{+}$ and if the threshold exceedance is positive or negative. However, what is universally true for the results in Fig. \ref{fig.SBextrem} is that despite $\vert S_{12}^{'\,C}\vert$ dominating the non-normal strain rate tensor, it is the axial terms that drive the local dynamics captured in the normal part of the strain rate tensor, although the topology varies from tube-like to two-dimensional to disc-like straining at different heights and for positive and negative threshold exceedances. 

The interaction between the local part of the straining and the non-local part of the vorticity captured in $\overline{\omega_{3}^{'\,C}S_{33}^{'\,B}\omega_{3}^{'\,C}}$ implicitly captures an important part of the model of \cite{motoori19} outlined in the introduction where small vortices are oriented at ninety degrees to larger structures (see also \cite{moffatt94}). Motoori and Goto stated (p.1103) that 

\textit{''$\ldots$large-scale vortices are stretched and created by the mean flow and they tend to align with the mean shear$\ldots$, whereas small-scale vortices apart from the wall are mainly stretched by the strain-rate field induced by large-scale vortices.''}

and their Figure 14 provides an example of such a phenomenon arising in the near-wall region. A non-local shearing in the plane of mean shear, i.e. $S_{12}^{'\,C}$ will, unless the tensor is Hermitian, result in a significant value for $\omega_{3}^{'\,C}$. Our results have indicated that this then causes a local response in the $S_{33}^{'\,B}$ term, i.e. orthogonal to the plane of shearing, which is at a smaller scale. That this terms plays a key role in near-wall enstrophy production then drives vortex development. Hence, not only do we confirm the importance of the fluctuating enstrophy production for the enstrophy production budget, we can show how the orthogonal orientation of larger and smaller structures can arise, while also explaining the double-peaked nature of the vertical profile. 

Our results have implications for near-wall control because, given the key role played by $\omega_{3}^{'\,C}$ for enstrophy production, if one wishes to re-laminarize a flow by damping vortex formation, the intervention must be sufficiently large in scale to perturb not only the small scale straining motions, but the larger scale vortical motions that interact with this straining term.  

\acknowledgments{The author was funded by a Japan Society for the Promotion of Science Bridge Fellowship and Leverhulme International Fellowship 2023-014. He is extremely grateful to Prof. Susumu Goto for hosting him in Osaka and to Prof. Goto and Dr Yutaro Motoori for discussions about this work.}

\bibliography{JFMk} 

\end{document}